\documentclass[twocolumn]{article}
\usepackage{amsmath}
\usepackage{amssymb}
\usepackage{fullpage}
\usepackage{graphicx}
\usepackage{amsthm}

\usepackage{algpseudocode}
\usepackage{algorithm}

\newcommand{\R}{{\mathbb R}}

\newcommand{\tr}{\operatorname{tr}}

\newtheorem{lemma}{Lemma}

\title{Even Simpler Deterministic Matrix Sketching}
\author{Edo Liberty\\ \it{Pinecone}}
\date{\nonumber}
 
\begin{document}
\maketitle
\thispagestyle{empty} 

\begin{abstract}
This paper provides a one-line proof of Frequent Directions (FD) for sketching streams of matrices. 
It simplifies the main results in \cite{Liberty13} and \cite{GhashamiP14}.
The simpler proof arises from sketching the {\it covariance} of the stream of matrices rather than the stream itself.
\end{abstract}

\section*{Introduction}
Let $X_t \in \R^{d \times n_t}$ be a stream of matrices. 
Let $C = \sum_{t=1}^{T}X_t X_t^T \in \R^{d \times d}$ be their covariance matrix. 
Frequent Directions \cite{Liberty13} maintains a rank deficient approximate covariance matrix $\tilde{C}_t \in \R^{d \times d}$ using Algorithm~\ref{fd}.
Set $\tilde{C}_0 \in \R^{d \times d}$ to be the all zeros matrix. Then, at time $t= 1,\ldots,T$ compute $\tilde{C}_{t} = \textsc{Update}(\tilde{C}_{t-1},X_t, \ell)$.  

\begin{algorithm}
  \caption{Frequent Directions (FD) Update}\label{fd}
  \begin{algorithmic}[1]
    \Function{Update}{$\tilde C_{t-1}, X_t, \ell$}
    \State $U_t\Lambda_t U_t^T = \tilde C_{t-1} + X_tX_t^T$ 
    \State \Return $\tilde C_{t} = U \cdot \max(\Lambda- I\cdot \lambda^t_{\ell}, 0)\cdot  U^T$
    \EndFunction
  \end{algorithmic}
\end{algorithm}

Above, $U_t\Lambda_t U_t^T$ is the eigen-decomposition of $\tilde C_{t-1} + X_tX_t^T$ and $\lambda^t_\ell$ is the its $\ell$'th largest eigenvalue. 
Note that the rank of $\tilde{C}_t$ is at most $\ell-1$ for all $t$ by construction. 
It can therefore be stored in $O(d \ell)$ space. Assuming $n_t < \ell$, the update operation itself also consumes at most $O(d\ell)$ space.

\begin{lemma}[simplified from \cite{GhashamiP14} and \cite{Liberty13}]
Let $\tilde{C}$ denote the approximated covariance produced by FD and $\lambda_i$ be the eigenvalues of the exact covariance $C$ in descending order.
For any $\ell$ and simultaneously for all $k < \ell$ we have
$$
\|C - \tilde{C}\| \le \frac{1}{\ell-k}\sum_{i=k+1}^{d}\lambda_i
$$
\label{fde}
\end{lemma}

\section*{Short proof Lemma~\ref{fde}}
Define $\Delta_t = X_tX_t^T - \tilde{C}_t + \tilde{C}_{t-1}$. 
Then $\sum_{t=1}^T \Delta_t = \sum_{t=1}^T X_tX_t^T - \sum_{t=1}^T(\tilde{C}_t - \tilde{C}_{t-1}) = C - \tilde{C}$ where $\tilde{C}$ stands for $\tilde{C}_T$, the final sketch.
 
Moreover, note that the top $\ell$ eigenvalues of $\Delta_t$ are all equal to one another because $\Delta_t = U_t \cdot \min(\Lambda_t, I\cdot \lambda^t_{\ell})\cdot  U_t^T$.
As a result $\| \Delta_t \| < \frac{1}{\ell-k} \tr(\bar{P}_k \Delta_t \bar{P}_k)$ for any projection $\bar{P}_k$ having a null space of dimension at most $k$.
Specifically, this holds for $\bar{P}_k$ whose null space contains the eigenvectors of $C$ corresponding to its largest eigenvalues.

\begin{align*}
\|C - \tilde{C}\|  &= \| \sum_{t=1}^T   \Delta_t\| 
\le  \sum_{t=1}^T  \| \Delta_t\| \\
&\le \frac{1}{\ell-k} \tr\left(\bar{P}_k \left(\sum_{t=1}^T \Delta_t\right) \bar{P}_k\right) \\
&\le \frac{1}{\ell-k} \tr\left(\bar{P}_k C \bar{P}_k\right) = \frac{1}{\ell-k}\sum_{i=k+1}^{d}\lambda_i
\end{align*}
Here we used that $\tr (\bar{P}_k \tilde{C} \bar{P}_k ) \ge 0$ because $\tilde{C}$ (and therefore $\bar{P}_k \tilde{C} \bar{P}_k$) is positive semidefinite. 
This completes the proof.

\bibliographystyle{unsrt}
\bibliography{simplefd}

\end{document}